# Methods of Fitting experimental data in several amplifiers for single spot radiometer

Kwang Rim Choe[1], Myong Chol Pak[2]

[1]High-Tech Research and Development, **Kim Il Sung** University, Pyongyang, Democratic People's Republic of Korea

[2]Department of Physics, **Kim Il Sung** University, Pyongyang, Democratic People's Republic of Korea

Abstract

Single spot radiometers are widely used to estimate temperature of body heated by high temperature, specially blast furnace, glass fusion, boiler of power plant, and so on. In previous papers on single spot radiometer a heated body was deduced according to interpolation functions between heated body temperature and photocurrent, which had previously measured. However, fitting functions required for current-to-voltage converter, non-inverting amplifier, and logarithmic amplifier have never reported. Therefore, in this paper the fitting functions are reported that can obtain the relationship between radiance temperatures of a body and outputs of amplifiers in single spot radiometer.

1.  Introduction

Literature [1] describes on structure, operational principle, and calibration of single spot radiometer, and structure of calibration source, and law of thermal radiation and presents interpolation function of photocurrent. Literatures [2, 3] describe on interpolation function and uncertainties in single spot radiometer . Also Literatures [4 - 6] present the characteristics of photodiode and the change of output in various amplifiers. Literatures [7, 9, 12] demonstrate method that photocurrent is amplified by current-to-voltage converter. Literatures [8, 10, 11] present various amplifiers based on a operational amplifier. Literatures [13, 15] describe the inverting and non-inverting amplifiers. Literature [14] describes non-inverting amplifier. Literatures [16-18] describe the principle of log-amplifier.

However, there are not the literatures demonstrating the experimental data fitting in various amplifiers of a single spot radiometer. Therefore the methods are proposed that fit with experimental data for various amplifiers of single spot radiometer, that is, current-to-voltage converter, non-inverting amplifier, and log-amplifier. We proposed the experimental interpolations that change current into voltage in the interpolation of literature [1], and leaded fitting function in non-inverting amplifier according to the interpolation of literature [1] and open circuit voltage in literature [4-6]. Also, using the output characteristics in log-amplifier and the interpolation of literature [1], we leaded the fitting function in log-amplifier. Then verification of the functions is made through the experiments.

2. Amplifiers of single spot radiometer.
(1) Current-to-voltage converter
Fig.1 shows circuit of current-to-voltage converter

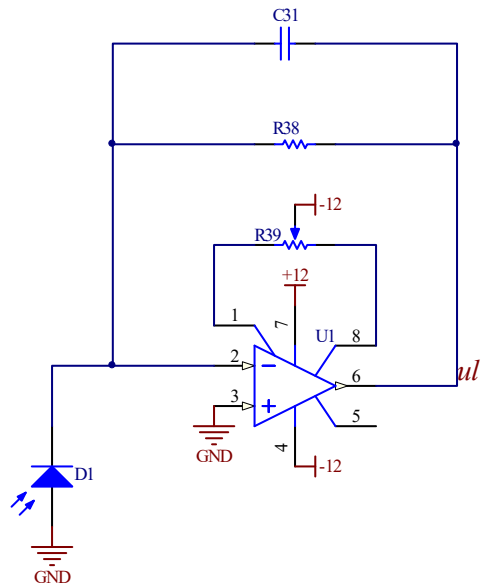

Fig.1 circuit of current-to-voltage converter

The photocurrent relates to the radiance temperature of target [1, 2],

$$i = \frac{C}{\exp\left(\frac{c_2}{AT+}\right)-1} \qquad (1)$$

Where $i$ is the photocurrent and A, B, and C are the constants obtained from experiment. $c_2$ is second radiation constant.
Also output voltage of current-to-voltage converter is as follow [4 - 6]

$$u_l = iR_{38} + u_0 \qquad (2)$$

According to equation (1) and (2), output voltage of current-to-voltage converter can write as follow.

$$u_l = \frac{CR_{38}}{\exp\left(\frac{c_2}{AT+B}\right)-1} + u_0 \tag{3}$$

(2) Non-inverting amplifier

Fig.2 shows non-inverting amplifier circuit. In non-inverting amplifier circuit, output voltage relates to input voltage as follow.

$$u_o = \left(1+\frac{R_{54}}{R_{55}}\right)V_{oc} \tag{4}$$

, where $u_o$ is output voltage and $V_{oc}$ the open circuit voltage of the photosensor, which expresses as follow[4-6].

$$V_{oc} = V_T \ln\left(1+\frac{i-I_{sh}}{I_{s0}}\right) \tag{5}$$

, where $I_{sh}$ is the current through the shunt resistance. $I_{s0}$ is the saturation current. $V_T$ is the thermal voltage.

Substituting the equation (1) into the equation (5), the relationship between $V_{oc}$ and radiance temperature $T$ is,

$$V_{oc} = V_T \ln\left(1+\frac{C/I_{s0}}{\exp\left(\frac{c_2}{AT+B}\right)-1} - \frac{I_{sh}}{I_{s0}}\right) \tag{6}$$

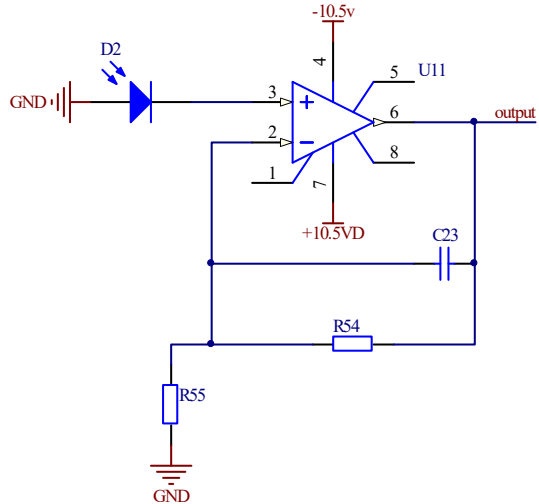

Fig.2 non-inverting amplifier circuit

Substituting equation (6) into equation (4), relation between output voltage and radiance temperature becomes,

$$u_o = \left(1 + \frac{R_{54}}{R_{55}}\right) V_T \ln\left(1 - \frac{I_{sh}}{I_{so}} + \frac{C/I_{so}}{\exp\left(\frac{c_2}{AT+B}\right)-1}\right) \tag{7}$$

When $\frac{C}{\exp\left(\frac{c_2}{AT+B}\right)-1} \ll I_0$ in equation (7), it become approximately,

$$u_o \approx \left(1 + \frac{R_{54}}{R_{55}}\right) V_T \frac{C/I_0}{\exp\left(\frac{c_2}{AT+B}\right)-1} - \frac{I_{sh}}{I_0} \tag{8}$$

When $\frac{C}{\exp\left(\frac{c_2}{AT+B}\right)-1} \gg I_0$ in equation (7), it become approximately,

$$u_o \approx \left(1 + \frac{R_{54}}{R_{55}}\right) V_T \left(-\frac{c_2}{AT+B} + \ln(C/I_0)\right) \tag{9}$$

Suppose that $\exp\left(\frac{c_2}{AT+B}\right) \gg 1$ and $\frac{I_{sh}}{I_0} \approx 0$ in equation (9). It can be seen from equation (8) and (9) that the radiance temperature relates nonlinearly to output voltage in low temperature region but linearly in high temperature region.

(3) logarithmic amplifier

Fig.3 shows logarithmic amplifying circuit. First stage is preamplifier in Fig.3. Preamplifier is current-to-voltage converter that photocurrent is converted to voltage. Output of preamplifier connects to input of logarithmic amplifier.

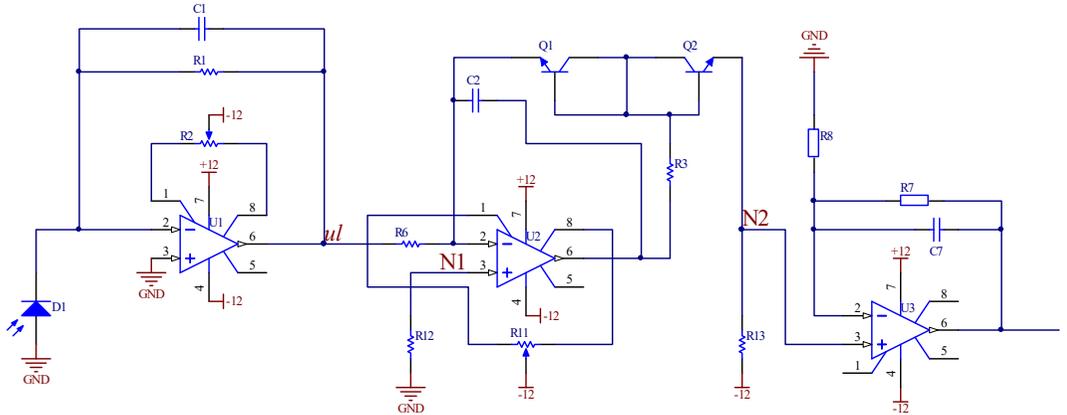

Fig.3 logarithmic amplifying circuit

Output signal of preamplifier become,

$$u_l = -(iR_1 + u_0) \tag{10}$$

In second logarithmic amplifier the current equation in node N1 become,

$$i_{c1} = i_{in} = \frac{u_l}{R_6} \approx I_{so1} e^{(V_{be1}/V_T)}$$

, where $i_{c1}$ the collector current of the transistor Q1, $I_{so1}$ the saturation current of the transistor Q1, and $V_{be}$ the base emitter voltage of the transistor Q1. From the equation above, $V_{be1}$ become

$$V_{be1} = V_T \ln \frac{u_l}{I_{so} \, R_6}$$

Otherwise the current equation in node N2 become

$$i_{c2} = I_{R13} \approx I_{so2} e^{(V_{be}/V_T)}$$

, where $i_{c2}$ the collector current of the transistor Q2, $I_{so2}$ the saturation current of the transistor Q2, and $V_{be2}$ the base emitter voltage of the transistor Q2. From the equation above, $V_{be2}$ become

$$V_{be2} = V_T \ln \frac{I_{R13}}{I_{so2}}$$

Since Q1 and Q2 are same in the characteristics, $I_{so1} = I_{so}$. Therefore the voltage $u_2$ in node N2 become

$$u_2 = V_{be} - V_{be2} = V_T \ln \frac{-u_l}{I_{R1} \, R_6} \qquad (11)$$

When the light is not incident on the photosensor, the output signal of amplifier U1 becomes nearly zero by adjusting resister R11. Then the voltage $u_2$ in node N2 becomes,

$$u_2 = V_T \ln \frac{-u_l}{I_{R13} R_6} - V_T \ln \frac{-u_1}{I_{R1} \, R_6} = V_T \ln \frac{iR_1 + u_0}{u_1} \qquad (12)$$

The last output voltage $u_3$ in the integrated circuit U1 becomes,

$$u_3 = \left(1 + \frac{R_7}{R_8}\right) V_T \ln \frac{iR_1 + u_0}{u_1} \qquad (13)$$

Substituting equation (1) into equation (13),

$$u_3 = \left(1 + \frac{R_7}{R_8}\right) V_T \ln \left(\frac{u_0}{u_1} + \frac{C}{\exp\left(\frac{C_2}{AT+B}\right) - 1} \frac{R_1}{u_1}\right) \qquad (14)$$

3. Experimental result
1) Current-to-voltage converter

Fig.4 shows experimental curve between radiance temperature and output signal in current-to-voltage converter. It can be seen from Fig.4 that radiance temperature of thermal emitting body relates logarithmically to the output voltage.

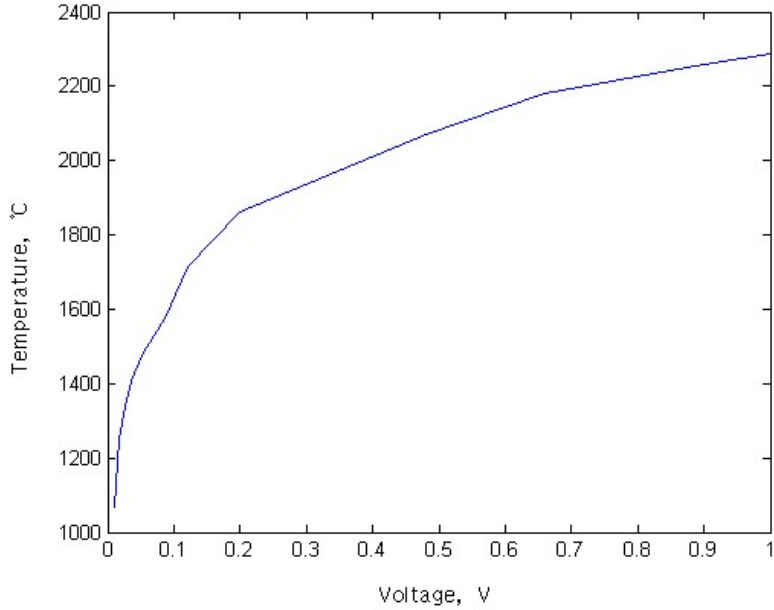

Fig.4 experimental curve between radiance temperature and output signal in current-to-voltage converter

From equation (3), fitting function become,

$$u_l = \frac{V_0}{\exp\left(\frac{c_2}{AT+B}\right)-1} + u_0 \tag{15}$$

, where $V_0 = CR_{38}$.

In Matlab, using fitting function (15), coefficients are deduced by non-linear least squares fitting. According to equation (15), the result is plotted in Fig.5. As the result, $A = 4.406e - 007$, $B = 0.0002766$, $V_0 = 2e + 007$, $u_0 = 16.14$.

Extended effective wavelength is [3],

$$\lambda_e = A + \frac{B}{T} \tag{16}$$

The extended effective wavelength calculated by equation (16) ranges 717.20~546.98 nm in temperature range of 1000~2600K.

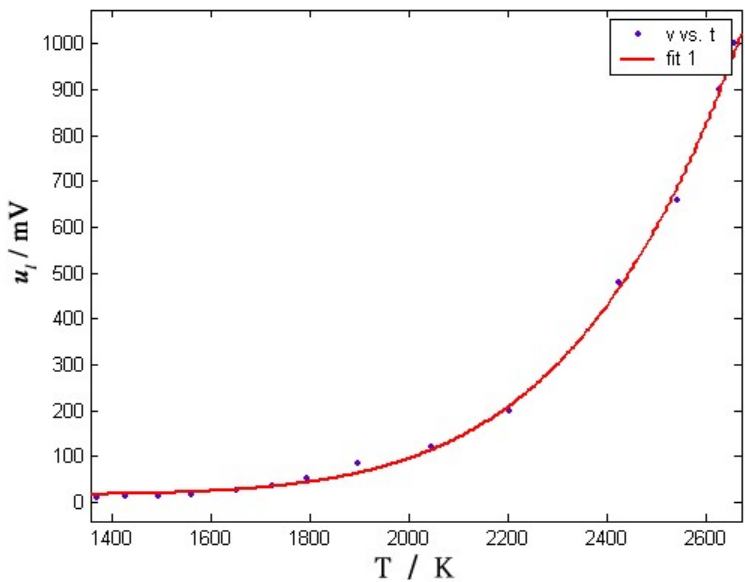

Fig.5 the result fitting by non-linear least squares

Calculating the photocurrent using $V_0 = CR_{38}$, $R_{38} = 1\ \text{M}\Omega$, and equation (1), that ranges 3.8770e-011 ~ 8.0777e-007A in temperature range of 1000~2600K.

2) Non-inverting amplifier

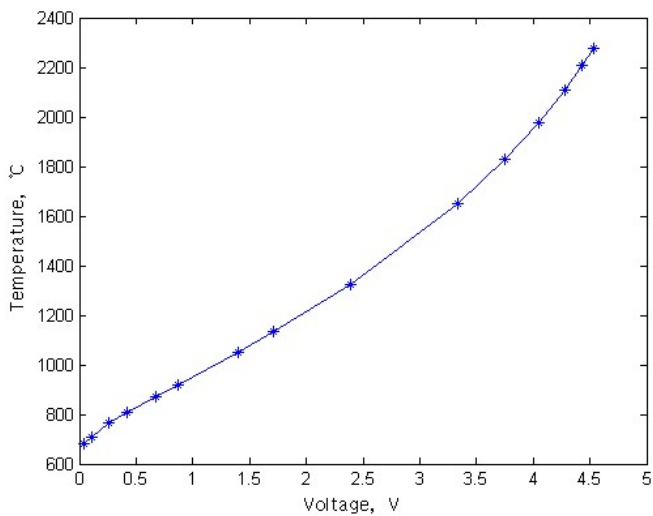

Fig.6 experimental curve between $u_o$ and $T$

Fig.6 shows experimental curve between $u_o$ and $T$ in non-inverting amplifier. As seen in Fig.6 $u_o$ relates nonlinearly to $T$ below 0.5V and $u_o$ is approximately proportional to $T$ over 0.5V. the reason of the experimental result is due to equation (8) and (9).

A/D value relates to $u_o$ as follow.

$$u_o = 5 \times AD/1023 \quad (17)$$

From equations (7) and (17), fitting function become,

$$AD = D \times \ln\left(1 + \frac{f}{\exp\left(\frac{C_2}{AT+B}\right)-1}\right) + g \quad (18)$$

, where $D = 1023 \times \left(1 + \frac{R_{54}}{R_{55}}\right)V_T/5$, $f = C/I_0$, and $g$ is a correct factor for voltage dropping by the filter. $\frac{I_{sh}}{I_0}$ is neglected in equation (18).

The fitting result is plotted in Fig.7.

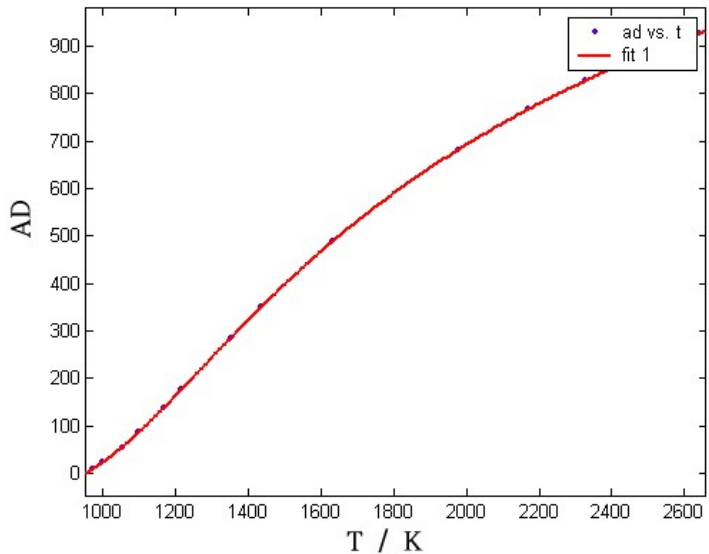

Fig.7 the result fitting by non-linear least squares

As the result, $A = 1.835e-006$, $B = 2.271e-005$, $D = 248.5$, $f = 958$, $g = -62.45$. The extended effective wavelength ranges 1.8577~1.8437 μm in temperature range of 1000~2600K.

3) Logarithmic amplifier

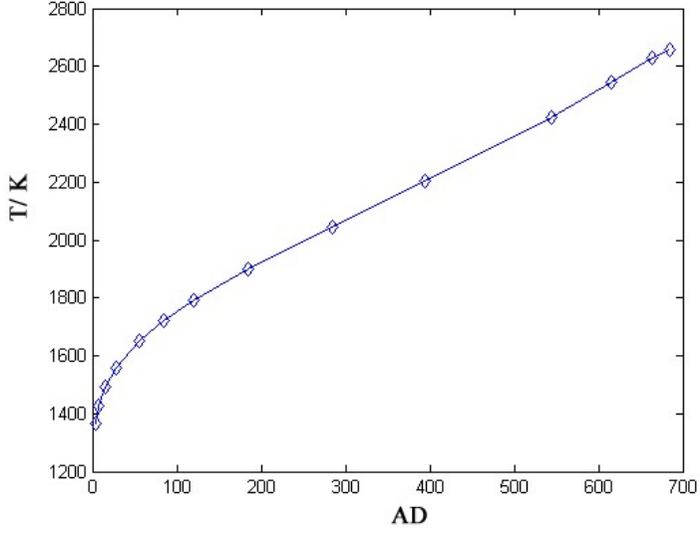

Fig.8 experimental curve between $u_3$ and $T$

Experimental curve between $u_3$ and $T$ is plotted on Fig.8. Optical system use same one as that in current-to-voltage converter.

Using equations (14) and (17), the fitting function becomes,

$$\mathrm{AD} = D \times \ln\left(e + \frac{f}{\exp\left(\frac{c_2}{AT+B}\right)-1}\right) + g \tag{19}$$

, where $D = 1023 \times \left(1 + \frac{R_7}{R_8}\right) V_T/5$, $f = \frac{CR_1}{u_1}$, $e = \frac{u_0}{u_1}$, and $g$ is a correct factor for voltage dropping by the filter. The fitting result is plotted in Fig.9.

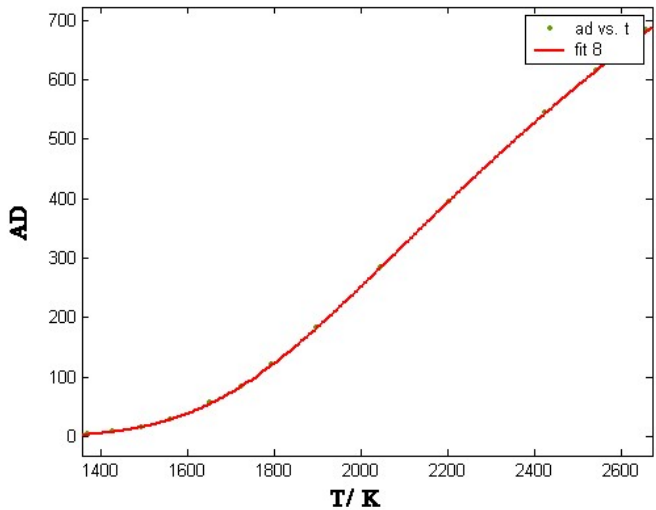

Fig.9 the result fitting by non-linear least squares

As the result, $A = 6.374e-007$, $B = 9.005e-005$, $D = 195$, $e = 0.9701$, $f = 1e+005$, $g = 2.337$. The extended effective wavelength ranges 727.45~672.03 ㎚ in temperature range of 1000~2600K. It is similar to that in current-to-voltage converter.

Conclusion

We establish new method, fitting experimental data from current-to-voltage converter, non-inverting amplifier, and log-amplifier for single spot radiometer through experiment in this paper.